\documentclass[conference]{IEEEtran}
\ifCLASSINFOpdf
  \usepackage[pdftex]{graphicx}
  \graphicspath{{./images/}}
  \DeclareGraphicsExtensions{.pdf,.jpeg,.png}
\else
\fi
\usepackage[tight,footnotesize]{subfigure}

\IEEEoverridecommandlockouts

\begin{document}
%
\title{Automatic Segmentation of Broadcast News Audio using Self Similarity Matrix}

\author{\IEEEauthorblockN{Sapna Soni\thanks{Sapna Soni is currently an
intern with the Speech and Natural Language Processing group at TCS Innovation Labs - Mumbai.}}
\IEEEauthorblockA{Institute of Technology,\\Nirma University\\
Ahmedabad, Gujarat- 380054\\
Email: 12mict41@nirmauni.ac.in}
\and
\IEEEauthorblockN{Imran Ahmed}
\IEEEauthorblockA{TCS Innovation labs - Mumbai,\\
Tata Consultancy Services Limited,\\
Yantra Park, Thane (West) - 400 601.\\
Email: ahmed.imran@tcs.com}
\and
\IEEEauthorblockN{Sunil Kumar Kopparapu}
\IEEEauthorblockA{TCS Innovation labs - Mumbai,\\
Tata Consultancy Services Limited,\\
Yantra Park, Thane (West) - 400 601.\\
Email: sunilkumar.kopparapu@tcs.com}}


%


\maketitle

\begin{abstract}

Generally audio news broadcast on radio is composed of music, commercials, 
news from correspondents and recorded statements in addition to the actual
news read by the  newsreader. 
When news transcripts are available, automatic segmentation
of audio news broadcast to time align the audio with the 
text transcription to build frugal speech corpora is essential. 
We address the problem of identifying segmentation in the
audio news broadcast corresponding to the news read by the newsreader so that
they can be mapped to the text transcripts.
The existing
techniques produce sub-optimal solutions when used to 
extract newsreader read segments.
In this paper, we propose a new technique which is able to identify
the acoustic change points reliably using an acoustic Self Similarity
Matrix (SSM). We describe the two pass technique in detail and 
verify its performance on real audio news broadcast
of All India Radio for different languages.

\end{abstract}


%
\IEEEpeerreviewmaketitle

\section{Introduction}

Audio segmentation is the process of partitioning an audio stream
into homogeneous segments. In other words it is the process of 
identifying the time instants in the audio stream when there is 
a change in the source/speaker. 
Audio and speaker segmentation are necessary pre-processing 
step for several important tasks such as automatic transcription, 
indexing, summarization, speaker diarization. Segmentation is applicable 
to  audio documents like 
broadcast news, telephone conversations, movies, etc. 
Speaker segmentation is the specific case of audio 
segmentation in which  segments corresponding to  the same speaker 
are identified. Speaker segmentation process generally  
involves the task of identifying and discarding non-speech regions in the 
audio; for instance silence, music, room noise, background noise or cross-talk.
Several algorithms have been proposed in literature for automatic 
segmentation and diarization of audio data in general; and speaker 
segmentation in particular. An overview of different techniques 
is discussed in \cite{reynolds} and 
\cite{xavier}. It should be noted that the techniques proposed 
in literature for audio segmentation are often tuned based in the 
type of data that they need to work on. 
For example, telephone conversations are spontaneous 
and contain frequent changes from one speaker to another
while the broadcast news audio data the change in the speaker in relatively
infrequent. As a result,
techniques for segmentation of broadcast news audio 
may not work for segmentation of telephone conversations \cite{imran2009} as
well.

In this paper we address the problem of segmentation of 
broadcast news audio, with the explicit task of extracting audio segments 
corresponding to the newsreader (or news anchor) speech. 
Generally, broadcast news audio is composed of music, 
commercials (advertisements), news from correspondents 
(reporters) and recorded statements in addition to the 
newsreader spoken speech. 
Automatic extraction of the broadcast news segment corresponding to the 
newsreader spoken speech becomes essential  
for automatic time alignment of speech and the available 
(newsreader) text transcription. Automatic segmentation of broadcast news if 
of particular use in development of a frugal speech corpora 
using broadcast news audio
(see for example \cite{frugalpaper,imran2013}).
Techniques proposed in literature tend to face difficulties 
in the form of sub-optimal solutions, 
when used on the task of automatic segmentation and extraction 
of the newsreader segments.
In this paper, we propose a novel technique which is able to 
identify the acoustic change points in an audio stream 
using an acoustic Self Similarity 
Matrix (SSM). SSM has been applied in music summarization 
and retrieval \cite{fxpal1,fxpal2} and also for audio 
segmentation \cite{fxpal3}, however these for a short duration audio. 
In broadcast audio news segmentation,  
the duration of the broadcast news audio is often longer than $10$ mins and
this results in a huge SSM, typically of the order $[60000 \times 60000]$
(assuming that we have a frame of length $10$ ms) making it infeasible to
extract homogeneous segments.
In order to avoid this kind of computation we propose 
the use of a different similarity measure.
We propose a two pass technique in which the long audio stream
is first divided into non-overlapping segments of $5$ seconds 
duration; and the MFCC feature vectors, extracted for every $10$ ms frame
is modelled as a multi-dimensional Gaussian distribution.
A SSM is then computed by calculating a pairwise similarity 
measurement between each of these segments using 
Bayesian Information Criterion (BIC) \cite{chen}.
The first level, coarse acoustic change points are obtained from this SSM; 
and then in the second pass the exact change point 
is obtained by computing a sliding window similarity 
measurement between overlapping segments of MFCC feature 
vectors around the identified coarse change point. 
We present the performance of this algorithm on 
broadcast news, in different Indian languages,
 available from All India Radio (AIR). 
AIR \cite{air} provides access to archives of 
news audio in several Indian languages, along with the transcripts corresponding
to the newsreader spoken speech.
The rest of the paper is organized as follows: Section \ref{sec:lit} we briefly
touch upon the issues with techniques proposed in the literature; we describe
the use of SSM in Section \ref{sec:ssm}. We describe our approach in Section
\ref{sec:our_approach}. Experimental results are described in Section
\ref{sec:experimental_results} and we conclude in Section
\ref{sec:conclusions}.

\section{Problem with Traditional Speaker Segmentation Techniques}
\label{sec:lit}
\begin{figure}
\centering
\includegraphics[width=0.450\textwidth]{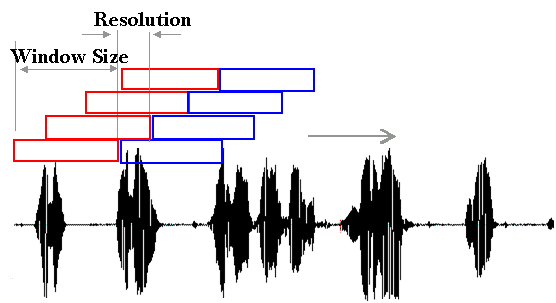}
\caption{Sliding window comparison technique for Speaker Segmentation.}
\label{fig:slidwin}
\end{figure}

\begin{figure}
\centering
\subfigure[Change detection on audio news (5 change points)]{
	\includegraphics[width=0.350\textwidth]{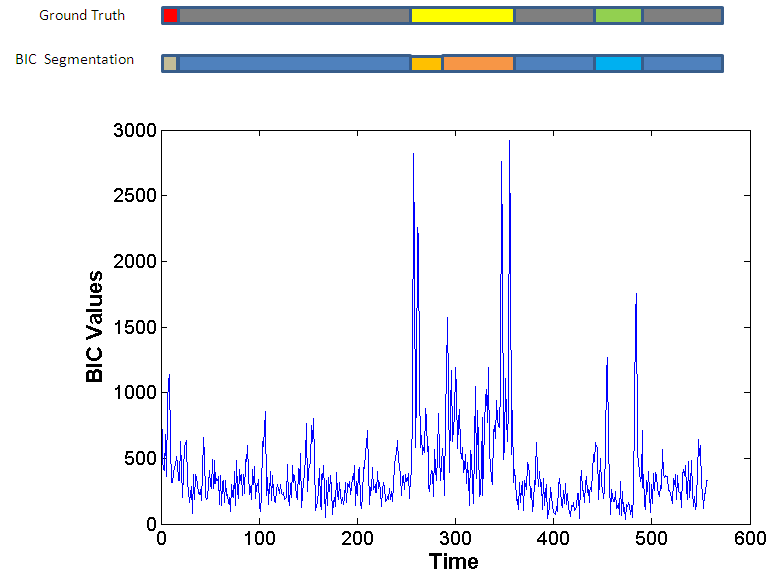}
	\label{fig:cmpgraph1}
}
\subfigure[Change detection on audio news (6 change points)]{
	\includegraphics[width=0.350\textwidth]{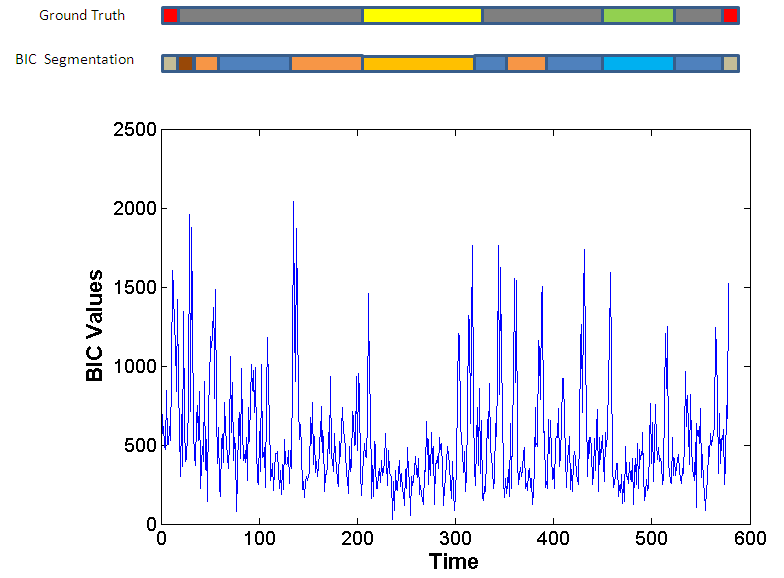}
	\label{fig:cmpgraph2}
}
\caption[]{Acoustic change detection using BIC;
the colored rectangles show the ground truth and the BIC based 
speaker segmentation obtained using LIUM.}
\label{fig:cmpgraph}
\end{figure}

Most of the 
proposed systems as described in \cite{reynolds,xavier} consist of
acoustic change detection (ACD), also called speaker change detection. 
ACD processes acoustic properties of an audio stream to identify 
instants of speaker change. The most common approach for 
change detection is to first divide the long audio steam 
into large number of smaller overlapping segments. Each small segment 
is represented by a set of speech feature vectors (MFCC or LSP)
and then a similarity metric (BIC or KL divergence) is 
calculated between any two adjacent speech segments (see  Figure
\ref{fig:slidwin}). 
Higher the dissimilarity, the more probability that the adjacent segments
belong to different speakers and hence a point of acoustic change. 
Typically, a chosen threshold value decides  whether the segments 
originate from the same or a different source. 
Figure \ref{fig:cmpgraph} shows the graph of the 
BIC (similarity) values computed every $0.1$ second, 
between adjacent segments of $2$ sec duration for 
two different broadcast news audio from AIR.
It also includes the ground truth for segmentation, 
and the segmentation obtained using the LIUM - 
an open-source state-of-the-art toolbox for Broadcast 
News Diarization \cite{lium}.
It can be observed  that
while BIC based technique works very well 
for one of the news audio (Figure \ref{fig:cmpgraph} (a)), 
it produces 
over-segmentation as seen for news audio in Figure \ref{fig:cmpgraph} (b).
While popularly used, BIC based technique
in literature tends to face (thresholding) 
problems when used on the task of automatic 
segmentation of the newsreader segments producing inconsistent results.
We propose a new technique which identifies the 
acoustic change points using an acoustic Self Similarity 
Matrix (SSM) next.

\section{Self Similarity Matrix (SSM) in Literature}
\label{sec:ssm}
SSM is a 2D characterization of all pairwise 
similarity measurements. It has been also referred to as 
{\em recurrence plots} or {\em dotplots}. SSM has been used 
in several applications including analysis of protein sequences,
visualizing structure of large text corpora, detecting periodic 
motion in video, music segmentation and 
summarization 
(see \cite{fxpal} and \cite{rp}). 
SSM has been applied for music summarization and retrieval 
\cite{fxpal1,fxpal2} and also for audio segmentation \cite{fxpal3}.
Here music files were represented as a sequence of 
feature vectors and a pairwise similarity measurement between each 
of these feature vectors is computed to construct a SSM.
In \cite{fxpal1} the feature vectors were calculated from the 
audio signal at a frame rate of $20$ frames per second (fps) and the typical 
length of the audio was around $200$ seconds; whereas 
in \cite{fxpal2} the feature vectors were calculated from the 
audio signal at a frame rate of $125$ fps and the 
typical length of the audio was around $10$ seconds.
This results in the size of the SSM of $[400 \times 400]$ and $[1250 \times 1250]$
respectively. 
However, in the task of broadcast news audio segmentation, the 
broadcast news is of length more than $10$ mins which results in a SSM of
size $[60000 \times 60000]$ 
(since our 
task is to identify newsreader read speech, we consider a $10$ msec audio
segment to compute the feature vector, which results in $100$ fps).
In order to avoid this kind of computation we propose use 
a different type of similarity measurement.

\section{Proposed Technique for Audio Segmentation}
\label{sec:our_approach}
We propose a two pass technique for segmentation of broadcast 
news audio, in order to extract 
the audio segments corresponding to the newsreader 
(or news anchor). The first pass we identify the coarse 
change points in the audio stream using SSM calculated over 
segments of duration $2-5$ seconds and in the second pass we 
identify the exact change point using SSM calculated for each 
MFCC frame ($10$ msec) in the region around the (coarse) change 
points detected in the first pass.

\subsection{First pass SSM}
\label{sec:1pass}
\begin{figure}[!t]
\centering
\includegraphics[width=0.350\textwidth]{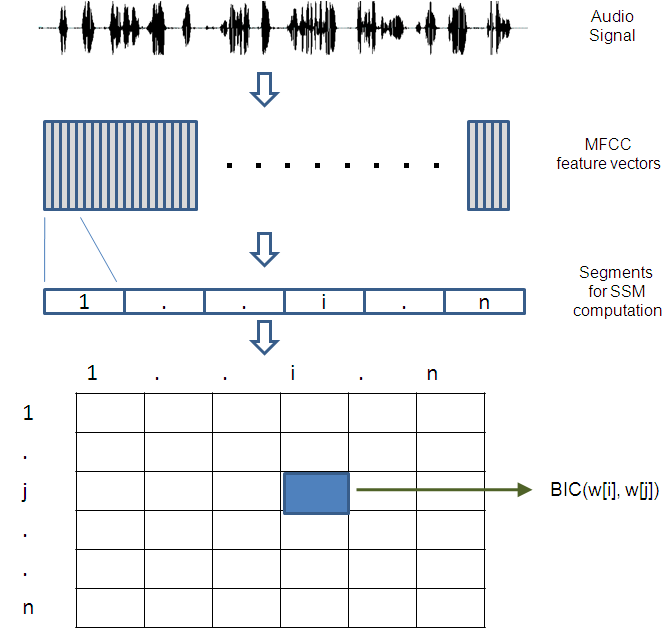}
\caption{Proposed technique 
for broadcast news segmentation.}
\label{ssmcomp}
\end{figure}

\begin{figure}[ht]
\centering
\subfigure[First pass SSM visualisation for audio in Figure \ref{fig:cmpgraph1}]{
	\includegraphics[width=0.350\textwidth]{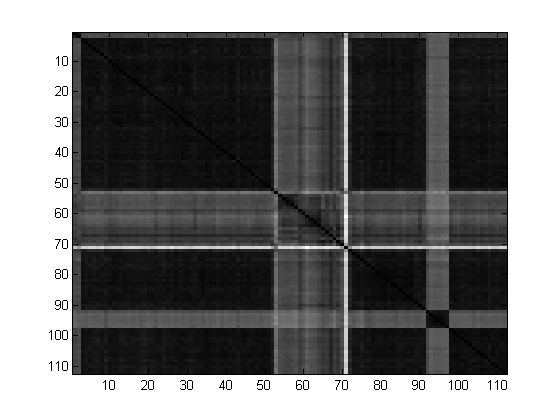}
	\label{fig:ssm1}
}
\subfigure[First pass SSM visualisation for audio in Figure \ref{fig:cmpgraph2}]{
	\includegraphics[width=0.350\textwidth]{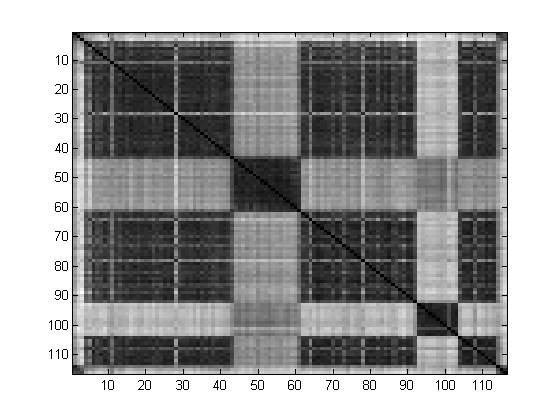}
	\label{fig:ssm2}
}
\caption[]{First pass SSM visualisation for audio in Figure \ref{fig:cmpgraph}}
\label{fig:ssm}
\end{figure}

Figure \ref{ssmcomp} depicts 
the steps used for computing the SSM in the 
first pass. As seen, the audio signal 
is converted into a sequence of MFCC feature 
vectors (on $10$ msec frames). 
These feature vectors are then combined 
together into groups of $2-5$ seconds. 
The grouped MFCC feature vectors in each of these 
segments are modelled as a multi-dimensional Gaussian distribution.
A SSM is then computed by calculating a pairwise similarity 
measurement between each of these segments using 
Bayesian Information Criterion \cite{chen}.
BIC is extensively used in speaker segmentation and 
clustering metric due to its simplicity and efficiency. 
The BIC similarity measure between the $i^{th}$ window 
($w_i$) and the $j^{th}$ window ($w_j$)  is given by 
\[
BIC(i,j) = \frac{N_{W}}{2}\log\left | \Sigma_{W}  \right | -
\frac{N_{w_i}}{2}\log\left | \Sigma_{w_i}  \right  | 
 - \frac{N_{w_j}}{2}\log\left | \Sigma_{w_j}  \right |,
\]
where, $\Sigma_{w_i}$, $\Sigma_{w_j}$ are respectively 
the co-variance matrices of the feature vectors in $w_i$ 
and $w_j$, and  $\Sigma_{W}$ is the co-variance matrix 
of all the feature vectors combined in the two windows $w_i$ and $w_j$
and ${N_{w_i}}/{2}$, ${N_{w_j}}/{2}$ and ${N_{W}}/{2}$ 
are respectively the number of feature vectors in the windows  
$w_i$, $w_j$ and $W$.
The BIC measure is an estimate of the measure of similar between two segments; 
Larger values of BIC, are an indication of dissimilarity between the two 
segments. Figure \ref{fig:ssm} shows a 
visualisation of the SSM for the two news audio 
shown in Figure \ref{fig:cmpgraph}. 
The darker the regions in these SSM's the higher 
the similarity of the corresponding audio segments.
It can be seen from Figure \ref{fig:ssm} that 
as we move along the SSM in the horizontal direction 
the (image) edges correspond to the change in darkness 
represent the acoustic change points in the audio.
As shown in Figure \ref{fig:ssm}, it is evident 
 that the proposed technique can reliably detect the change points in the 
audio stream. However, it should be noted that localization precision 
of the detected change point using the SSM in the first pass 
directly proportional to the length of the segment. Hence, the change 
points obtained in this pass ($2-5$ seconds) are coarse. We 
perform a second pass to localize and detect the actual change point.

\subsection{Second Pass}
\label{sec:2pass}
In this pass a segment ($10-20$ second) is taken around the 
change point detected in the first pass and the exact change 
point is obtained by computing a sliding window similarity 
measurement between overlapping segments of MFCC features
in a manner to the technique discussed in Section \ref{sec:lit}. 
In this case it is already known that there is one change 
point in the entire segment and hence the highest peak in 
the change detection graph can be reliably selected as the 
change point; without having to worry about an appropriate choice of 
a threshold value.
Once the exact acoustic change points in the audio are identified 
the longest segment is treated as the newsreader and 
other segments having a low BIC similarity with this 
segment are treated as the segments of the newsreader. 

\section{Experiments for Performance Analysis}
\label{sec:experimental_results}

In order to evaluate the performance of the proposed 
algorithm we used $10$ broadcast news sourced from 
\cite{air}. Each of these audio news is of 
$10$ mins duration and consists of audio segments from the 
newsreader, news correspondent or reporter, commercials 
and music. Each of these audio files is processed to 
extract MFCC feature vectors, from $25$ msec frames of 
speech, every $10$ msec. In order to perform the first 
pass the MFCC feature vectors for each audio are grouped 
into a segment of $5$ second duration and a SSM is computed 
as discussed in Section \ref{sec:1pass}. This gives the 
coarse change points in the audio. Then as discussed 
in Section \ref{sec:2pass} a $20$ second segment is taken 
around the coarse change point of first pass. 
A sliding window BIC comparison is carried for $2$ seconds
of two adjacent windows, every $100$ msec. The peak of the 
resulting BIC change detection graph is chosen as the 
exact change point. For the purpose of comparison 
the number of segments extracted using the proposed 
two pass algorithm is compared with the actual number of 
segments in the audio and also to the number of segments
obtained using the LIUM open-source toolbox for Broadcast 
News Diarization \cite{lium}.

\begin{figure}
\centering
\includegraphics[width=0.350\textwidth]{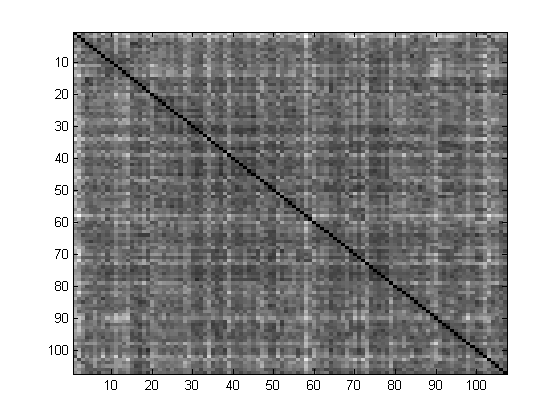}
\caption{SSM visualisation for broadcast news audio without any change points.}
\label{fig:nochg}
\end{figure}

Table \ref{tab:exp} shows the results for each of the $10$ news 
audio. It can be seen that the 
proposed algorithm detects almost the exact number of 
segments whereas the LIUM system tends to over-segment 
the audio files. 
It should be noted that two of the $10$ audio news 
consist of only the newsreader and the proposed 
algorithm does not show any change points in these
audio files. This can also be seen through the SSM of
one these audio files, as shown in Figure \ref{fig:nochg}.
It should also be noted that in both the 
cases the actual change points were accurately marked. 
However, the LIUM system faces the problem of selecting 
change points as discussed in Section \ref{sec:lit}.
Whereas the proposed algorithm is able to avoid this 
using the SSM in the first pass.

\begin{table}[h]
\renewcommand{\arraystretch}{1.3}
\caption{Comparison of Number of Segments generated}
\label{tab:exp}
\begin{center}
\begin{tabular}{|l||c|c|c| }
\hline 
{\bf News Broadcast} & {\bf Actual} &  {\bf by LIUM} & {\bf by SSM} \\
\hline
Hindi\_Patna1Oct & 8 & 15 & 7 \\
\hline
Hindi\_Indore11Oct & 7 & 21 & 7 \\
\hline
Hindi\_Patna10Oct & 5 & 15 & 4 \\
\hline
Hindi\_Shimla3Feb & 7 & 6 &7 \\
\hline
Hindi\_Shimla20Feb & 3 & 12 & 3 \\
\hline
Telugu\_Gangtok3Oct & 1 & 1 & 1 \\
\hline
Telugu\_Vijaywada3Oct & 5 & 7 & 5 \\
\hline
Telugu\_Vijaywada28Sept & 5 & 4 & 5 \\
\hline
Telugu\_TeluguNSD27Sept & 1 & 7 & 1 \\
\hline
Telugu\_Hyderabad27Sept & 3 & 3 & 3 \\
\hline
{\bf Total} & {\bf 45} & {\bf 91} & {\bf 43} \\
\hline
\end{tabular}
\end {center}
\end {table} 

\section{Conclusion}
\label{sec:conclusions}
In this paper we address the problem of segmentation of 
long audio stream like broadcast news audio, 
in order to extract the audio segments 
corresponding to the newsreader. The existing systems 
and techniques proposed in literature give sub-optimal solutions
when used for the task of automatic segmentation and extraction 
of the newsreader segments from a long audio stream.
We proposed a two pass technique; in the first pass a number of 
coarse acoustic change points using an acoustic Self Similarity 
Matrix are identified; when in the second pass the exact 
position of the acoustic change points is found. The proposed algorithm 
is evaluated for its performance on broadcast news from All India Radio
and shows to perform with higher accuracy compared to other
available audio segmentation tools.

\bibliographystyle{IEEEtran}
\bibliography{references}

\end{document}